\documentclass{svjour3}   
\smartqed  
\usepackage[top=3cm, bottom=3cm, left=3cm, right=3cm]{geometry}
\usepackage{amssymb}
\usepackage{amsmath}
\usepackage{dsfont}
\usepackage{enumerate}
\usepackage{graphicx}
\usepackage{float}
\usepackage{subfig}
\usepackage{caption}
\usepackage{tabularx}
\usepackage{multirow}

\newcommand{\ket}[1]{\left|{#1}\right\rangle}
\newcommand{\bra}[1]{\left\langle{#1}\right|}

\begin{document}

\title{Data reconstruction based on quantum neural networks}

\author{Ming-Ming Wang \and
     Yi-Da Jiang}

\institute{Ming-Ming Wang \at
            Shaanxi Key Laboratory of Clothing Intelligence, School of Computer Science, Xi'an Polytechnic University, Xi'an 710048, China\\
            \email{bluess1982@126.com}\\
           Yi-Da Jiang \at
            Shaanxi Key Laboratory of Clothing Intelligence, School of Computer Science, Xi'an Polytechnic University, Xi'an 710048, China\\
}

\date{Received: date / Accepted: date}

\maketitle

\begin{abstract}
Reconstruction of large-sized data from small-sized ones is an important problem in information science, and a typical example is the image super-resolution reconstruction in computer vision. Combining machine learning and quantum computing, quantum machine learning has shown the ability to accelerate data processing and provides new methods for information processing.
In this paper, we propose two frameworks for data reconstruction based on quantum neural networks (QNNs) and quantum autoencoder (QAE). The effects of the two frameworks are evaluated by using the MNIST handwritten digits as datasets. Simulation results show that QNNs and QAE can work well for data reconstruction.
We also compare our results with classical super-resolution neural networks, and the results of one QNN are very close to classical ones.
\end{abstract}

\section{Introduction}

Advances in computer algorithms and hardware have made machine learning a great success in the fields of computer vision \cite{HeChen-35,LiuLin-34}, data mining \cite{WangCao-38}, medical diagnosis \cite{TanLiu-37}, cyber security \cite{LvSingh-39}, etc.
Deep neural networks \cite{GoodfellowBengio-312}, also known as multilayer perceptrons, play a vital role in machine learning for dealing with a vast majority of tasks such as classification, regression, and unsupervised learning.

With great achievements in quantum computation \cite{Shor97,Grover97,HHL-2009} and quantum communications \cite{GisinRibordy-370}, quantum machine learning (QML) \cite{SchuldSinayskiy-136,BiamonteWittek-87,DunjkoBriegel-42} has been developed for processing quantum data and accelerating classical machine learning algorithms.
As the counterpart of neural networks, quantum neural networks (QNNs) have also been studied \cite{Kak-174,Ronald-173,BehrmanNiemel-189,VenturaMartinez-190,MatsuiTakai-192,SchuldSinayskiy-194,BeerBondarenko-166}.
A typical QNN model is the parameterized quantum circuit (PQC) \cite{plesch2011quantum,schumacher1995quantum} which constructs a set of learnable parametric quantum gates whose learnability is dependent on classical parameters.
These parameters are optimized iteratively by minimizing a loss function.
Recently, a variety of QNNs have been proposed, such as quantum convolutional neural networks \cite{CongChoi-181}, quantum autoencoder (QAE) \cite{RomeroOlson-104,Bravo-Prieto-203}, quantum generative adversarial networks (QuGAN) \cite{LloydWeedbrook-180,BracciaCaruso-44}, and hybrid quantum networks \cite{Hellstem-47}, etc.

Data reconstruction is an important task in information processing. A typical example of data reconstruction is the image super-resolution reconstruction \cite{Irani-325,DongLoy-324}, which recovers a high-resolution image from a given low-resolution one. It is an important subject in computer vision, where the reconstructed high-resolution image can be used to solve problems such as image classification, semantic segmentation, etc.
Most of the existing studies for super-resolution reconstruction problems are solved by classical neural networks. In the field of quantum information, reconstructing high-dimensional data from low-dimensional ones is still an interesting problem.

Some research is related to the data reconstruction problem based on QML. For example, the compression of quantum data based on QAE has been studied theoretically  \cite{RomeroOlson-104,Bravo-Prieto-203} and experimentally \cite{PepperTischler-208,HuangMa-209}, where the decompression process is similar to the data reconstruction since they all try to recover large-sized data from small-sized ones.
Besides, QuQANs \cite{SteinBaheri-51,RudolphToussaint-52} have been applied to generate artificial quantum and classical data similar to real ones, but they have a different type of input compared to the data reconstruction task.

To deal with the data reconstruction problem in quantum field, we propose two data reconstruction frameworks based on QNNs and QAE in this paper.
The main contributions of this paper include (1) a QNN-based framework for data reconstruction; (2) a QAE-based framework for data reconstruction; (3) numerical simulations of two frameworks using MNIST handwritten digits.

The rest of the paper is organized as follows.
Some preliminaries of the paper are introduced in Sect. 2.
Two data reconstruction frameworks based on QNNs and QAE are presented in Sect. 3.
The numerical simulation of two frameworks on MNIST handwritten digits are shown in Sect. 4.
The paper is further discussed and concluded in Sect. 5.

\section{Preliminaries}

\subsection{Loss functions}

A loss function evaluates the difference between the output of the model and the target value. It defines the optimization direction of a machine learning algorithm. The choice of different loss functions often results in network models with different expressiveness.
$L_1$ and  $L_2$ loss functions are widely used in classical neural networks.
The $L_1$ loss \cite{SaraAkter-66}, also called the mean absolute error (MAE), is defined as
\begin{equation}\label{eq1}
    L_1 = \frac{1}{n} \sum_i^n  \mid y_i - f(x_i)  \mid,
\end{equation}
where $y_i$ is the target value of the reference and $ f(x_i)$ is the predicted value (output) of the model, and $n$ is the sample size.
While the $L_2$  loss function, also known as the squared error loss (MSE), is computed as
\begin{equation}\label{eq2}
    L_2 = \frac{1}{n} \sum_i^n  ( y_i - f(x_i))^2.
\end{equation}

For a QNN, suppose it performs a unitary operation $\mathcal{N}$ on the input state $\rho^{\text{in}}_i$ for generating the output state $\mathcal{N}(\rho^{\text{in}}_i)$, the loss function of the QNN can be defined as
\begin{equation}\label{eq3}
L_q=  \frac{1}{n}\sum_{i=1}^n (1 - F(\mathcal{N}(\rho^{\text{in}}_i),\sigma^{\text{ref}}_i)),
\end{equation}
where
$\sigma^{\text{ref}}_i$ denotes the target (reference) state, $F$ denoted the fidelity of two quantum states. The fidelity of two pure states can be written as \cite{NielsenChuang-404}
\begin{equation}
F (\ket{\phi}\ket{\psi})= |\left\langle{\phi}\right. \ket{\psi}|^2.
\end{equation}

\subsection{Amplitude encoding}

For encoding classical data into quantum states, a common way is to use the amplitude coding, which encodes an $N$-dimensional vector into $M$ qubits, where $M = \lceil \log_2 N \rceil$.
Given a vector $\vec{x}=(x_1, x_2, \cdots,  x_i, \cdots, x_N)$,  it needs to be normalized to $\vec{\tilde{x}}=(\tilde{x}_1, \tilde{x}_2, \cdots,  \tilde{x}_i, \cdots, \tilde{x}_N)$ with
\begin{equation*}
\tilde{x}_i = \frac{x_i}{\sqrt{\sum x_i^2 }}.
\end{equation*}

Then, $\vec{x}$ can be encoded as
\begin{equation*}
\sum_i^N \tilde{x}_i \ket{i}.
\end{equation*}
where $\ket{i}$ is the computational basis state.

\subsection{The QNN model}

\begin{figure}
\centering
\includegraphics [scale=0.5]{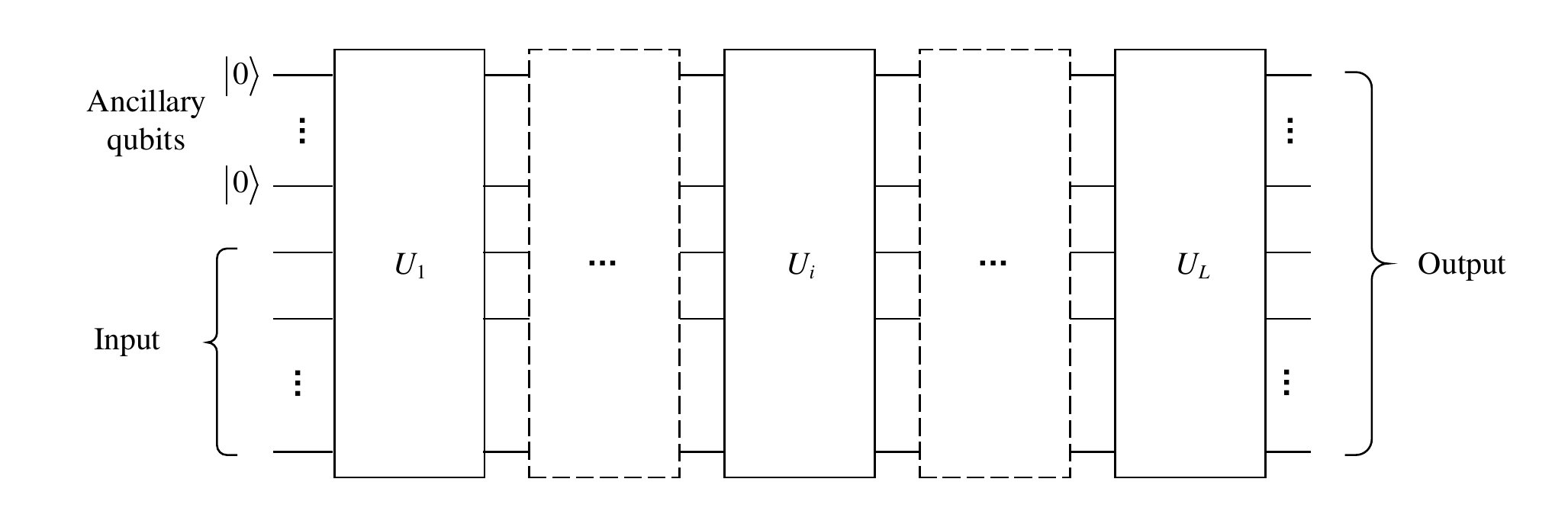}
  \caption{Data reconstruction based on parameterized QNNs.}
  \label{fig1}
\end{figure}

PQCs are used to construct QNNs for representing the target function of the learning task. That is, we define a quantum circuit that depends on a set of tunable parameters $\vec{\theta}$, which are encoded in the circuit for performing the unitary operation $U(\vec{\theta})$.
These parameters are updated iteratively according to the loss value during the training process. As is shown in Fig. \ref{fig1}, suppose a parameterized QNN are consisting of $L$ layers, the action of the QNN can be represented by a unitary matrix as $U(\vec{\theta})=U_1(\vec{\theta}_1)\cdots U_i(\vec{\theta}_i) \cdots U_L(\vec{\theta}_L)$.

\begin{figure}
\centering
\subfloat[Circuit 1]{\label{fig2-1}\includegraphics [scale=0.35]{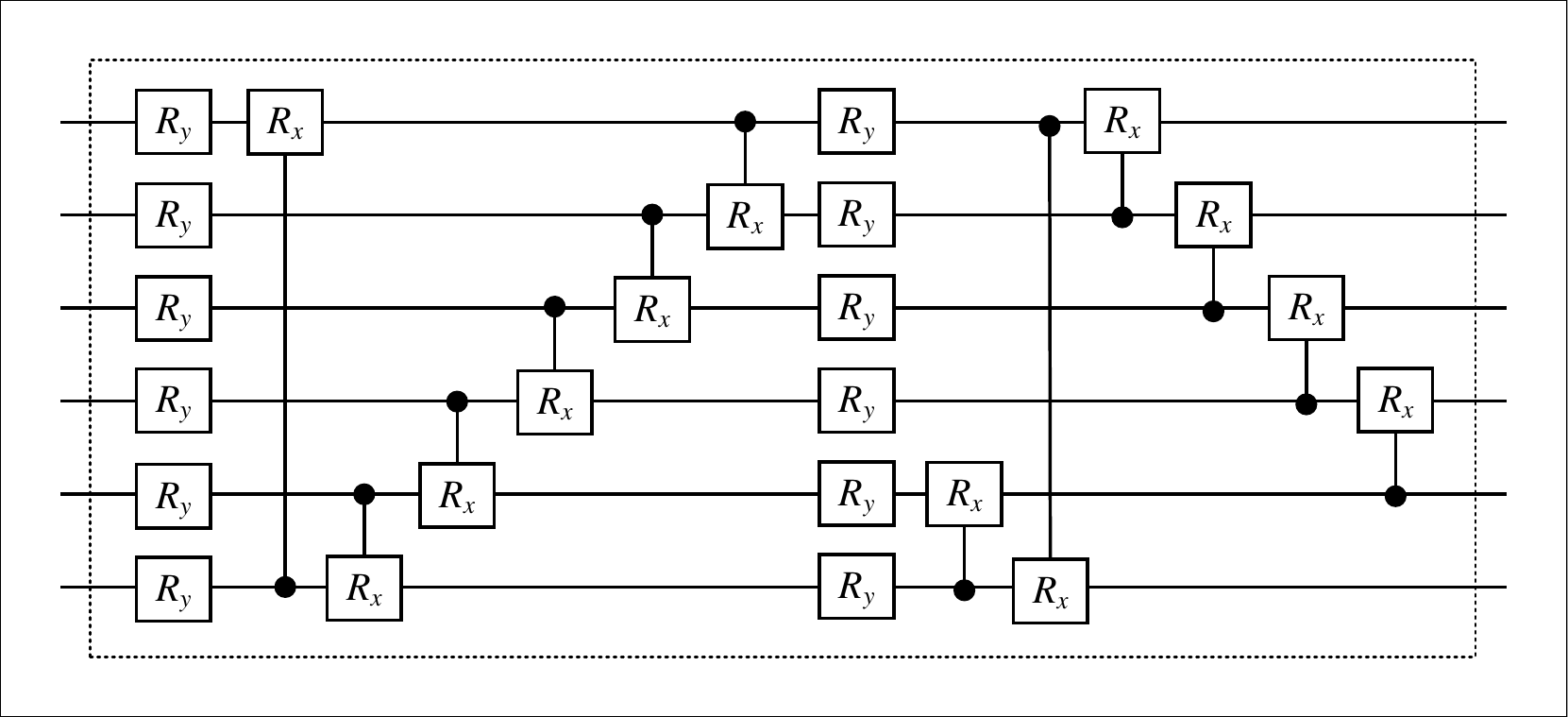}}
\subfloat[Circuit 2]{\label{fig2-2}\includegraphics [scale=0.35]{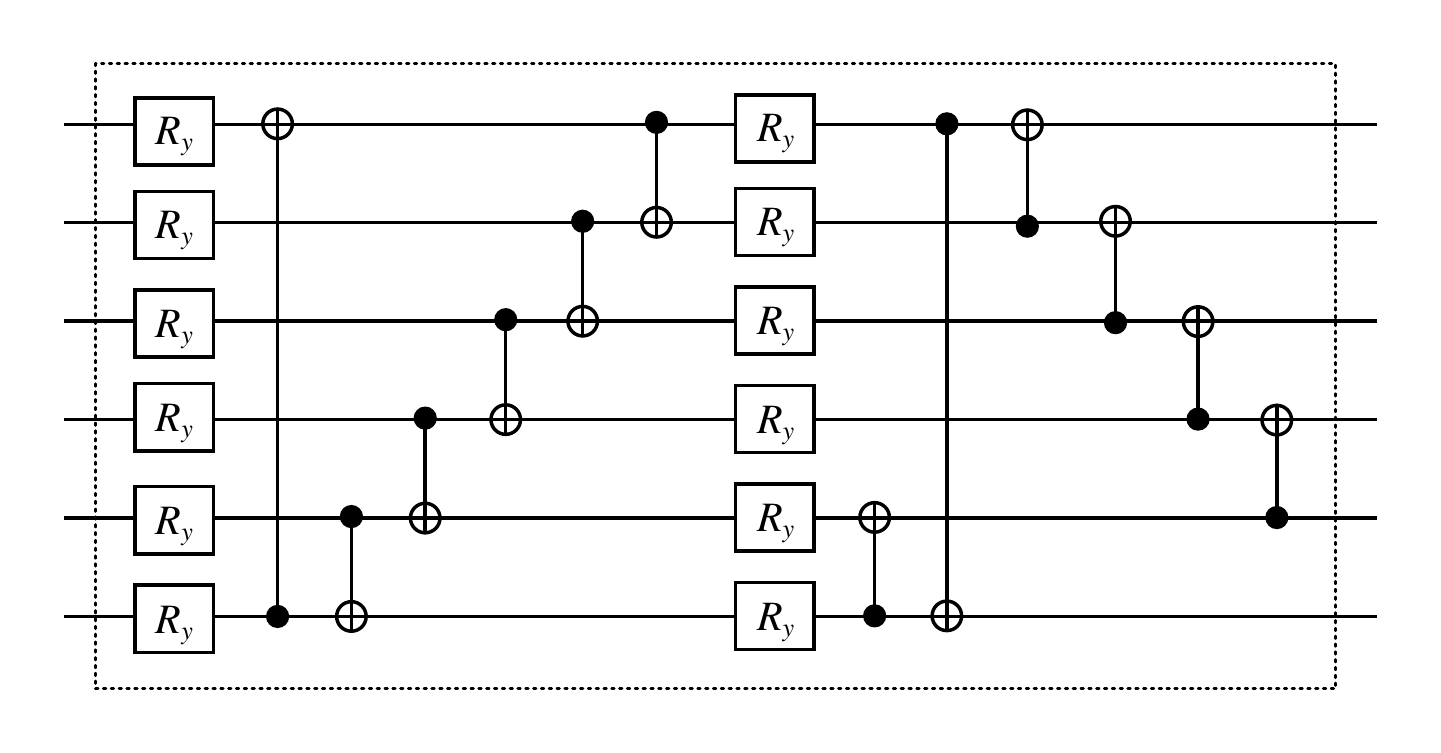}}
\subfloat[Circuit 3]{\label{fig2-3}\includegraphics [scale=0.35]{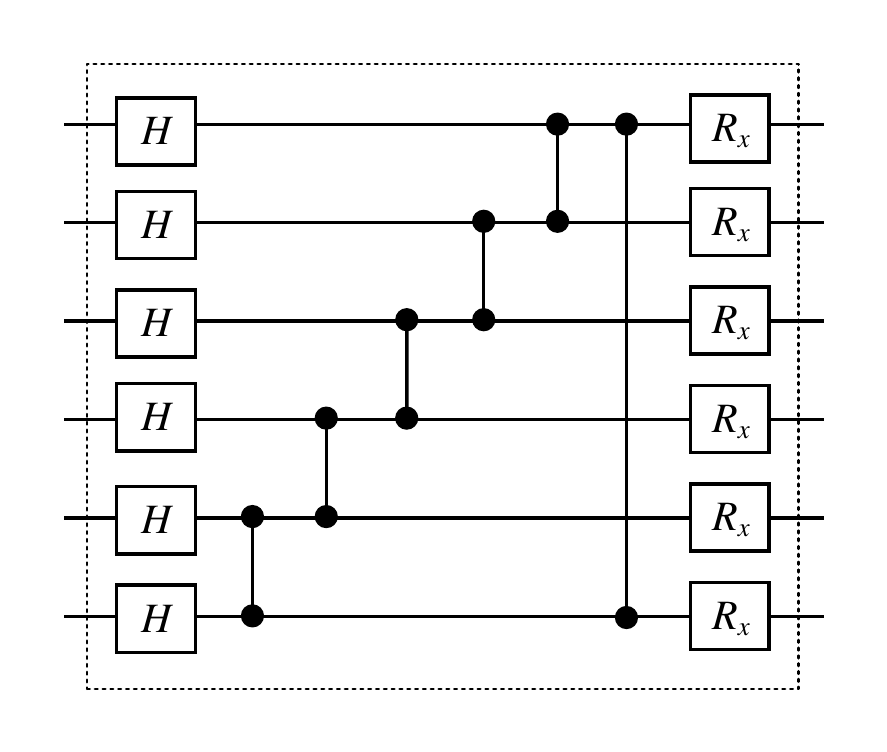}}
  \caption{Three circuits used for QNN as a layer $U_i$.}
\label{fig2}
\end{figure}

Each layer $U_i$ of the QNN is consisted of a set of one-qubit rotation gates (such as $R_x(\theta)$ and $R_y(\theta)$), and two-qubit controlled gates (such as CZ and CNOT gates) \cite{NielsenChuang-404,KrausCirac-68}.
Three circuits used for our QNNs are shown in Fig. \ref{fig2}, where Circuit 1, 2, and 3 are from Refs. \cite{SimJohnson-1380,SchuldBocharov-1381}.

\section{Data reconstruction frameworks}

Two frameworks are proposed to deal with the data reconstruction problem, i.e., the QNN-based framework and the modified QAE-based framework.

\subsection{The QNN-based framework}

\begin{figure}
\centering
\includegraphics [scale=0.5]{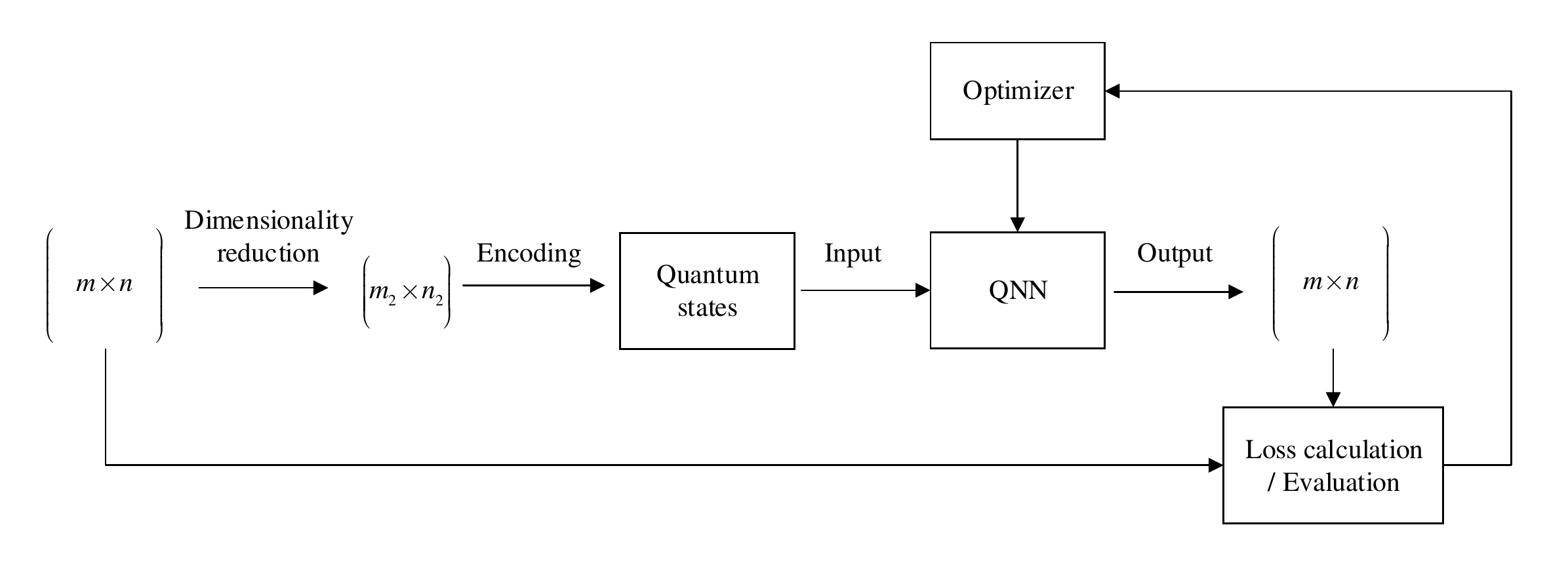}
  \caption{The QNN-based framework for data reconstruction.}
  \label{fig3}
\end{figure}

The framework of data reconstruction based on QNN is shown in Fig. \ref{fig3}.
For training a QNN, we start from the training data with dimension $m\times n$ and resample them into small-sized data with dimension $m_2\times n_2$ \cite{ParkerKenyon-70}. These smaller dimensional data are encoded into quantum states, which further fit into the QNN as the input states. Together with some auxiliary qubits, $U(\vec{\theta})$ is performed for generating the output states, which are decoded as the reconstructed data with dimension $m\times n$.
The loss function is calculated between the training data and the reconstructed data for optimizing $\vec{\theta}$ iteratively.

For testing the QNN, it performs only once to output the reconstructed data, while the effect of the QNN is measured by the average loss between the reference data and the reconstructed data.

\subsection{The QAE-based framework}

\begin{figure}
\centering
\includegraphics [scale=0.45]{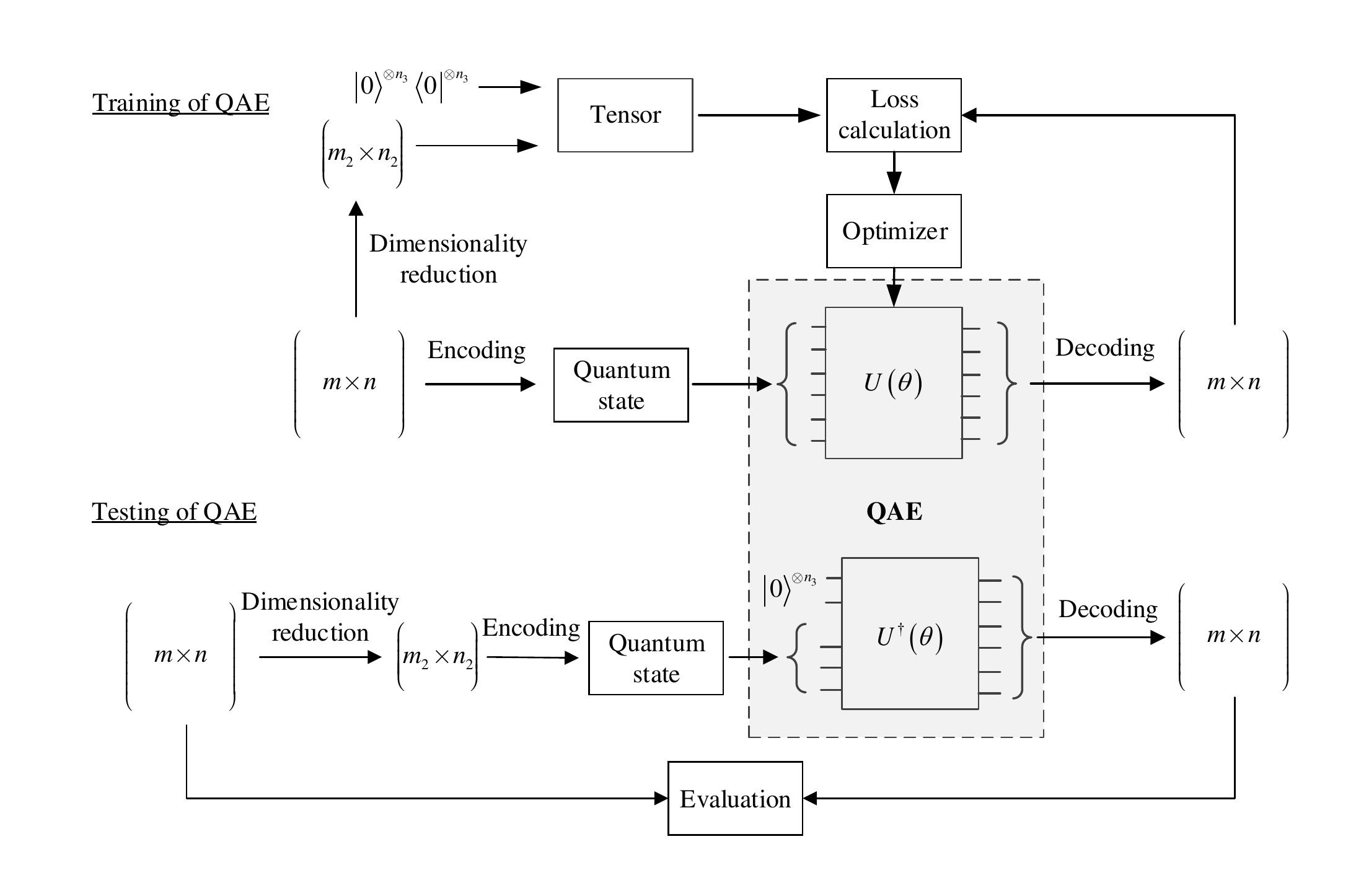}
  \caption{The framework of data reconstruction based on modified QAE.}
  \label{fig3-2}
\end{figure}

\begin{figure}
\centering
\includegraphics [scale=0.4]{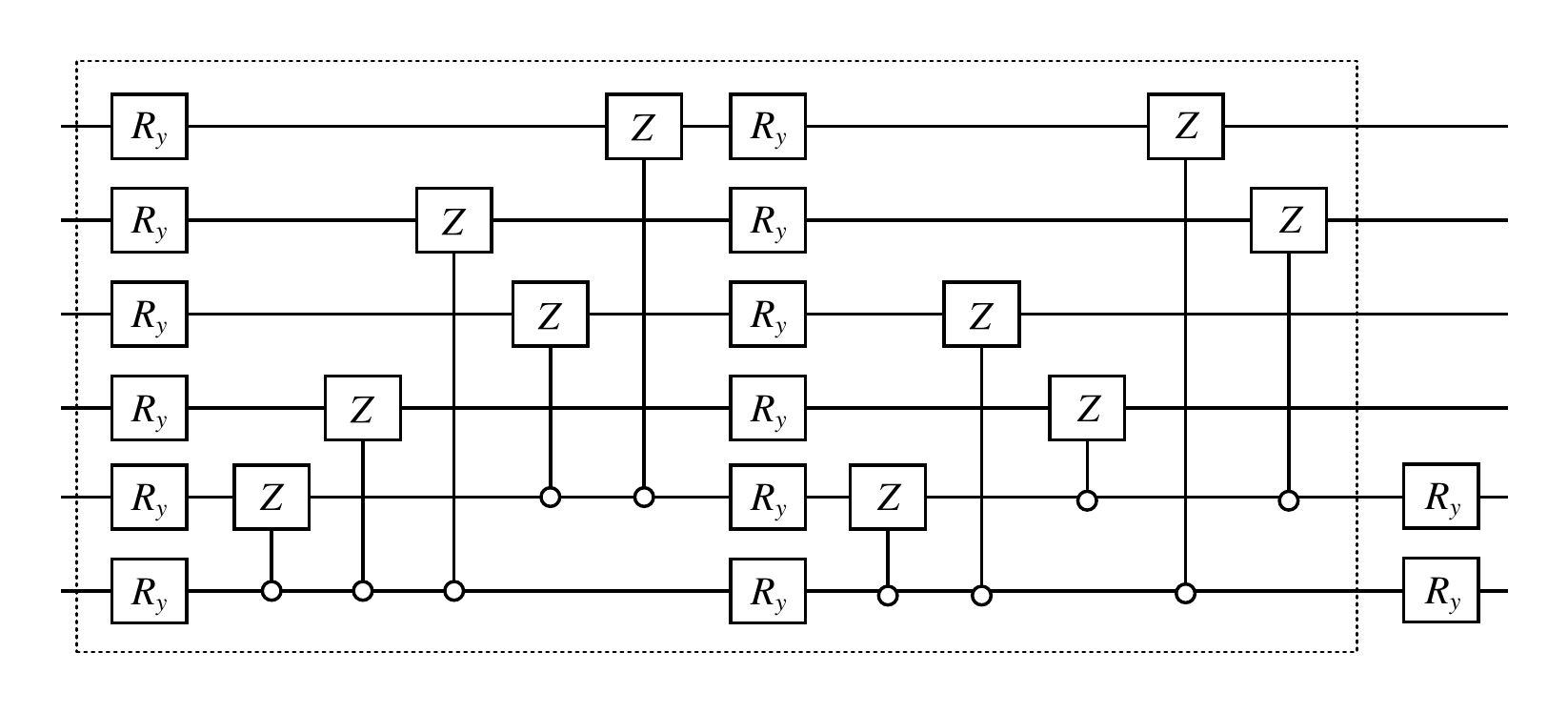}
\caption{The circuit used for QAE.}
\label{fig3-3}
\end{figure}

QAE \cite{RomeroOlson-104,Bravo-Prieto-203} has been used for the compression of quantum states. Here, we modify the QAE for dealing with the data reconstruction task, where the modified QAE-based framework is shown in Fig. \ref{fig3-2}.
Using the same training and testing sample, we first train the QAE represented by $U(\vec{\theta})$ using the method similar to Ref. \cite{Bravo-Prieto-203}, also see Fig. \ref{fig3-3}.
While $U^{\dag}(\vec{\theta})$ is applied as the data reconstruction circuit for evaluation of the QAE.

\section{Simulation}

To evaluate the effectiveness of QNN and QAE frameworks, we use the MNIST handwritten digits \cite{lecun1998gradient} from Scikit Learn as the dataset.
Three training sets are chosen here, i.e., the image 0 set, image 1 set, and the mixture of image 0 and image 1 set, where each set contains 50 images for training and 30 images for testing.

For training of each QNN shown in Fig. \ref{fig3}, an $8\times 8$ sized handwritten digit is downsampled into a $4\times 4$ sized image. Then, the $4\times 4$ sized image is encoded into a 4-qubit state by the amplitude encoding, which is the input of each QNN. With a 2-qubit auxiliary state, QNNs are implemented on a 6-qubits system.
After the operation of the QNN, the output 6-qubit state is decoded as an $8\times 8$ sized image for calculation of the loss.

While for the training of the QAE, an $8\times 8$ sized handwritten digit is encoded into a 6-qubit state as the input of QAE. And an $8\times 8$ sized image is decoded from the output of QAE. Besides, the $8\times 8$ sized handwritten digit is also downsampled into a $4\times 4$ sized image. Tensoring with $\ket{0}\bra{0}^{n_3\otimes n_3}$, the loss is calculated between these two matrices (see Fig. \ref{fig3-2}).

To compare QNNs and QAE with classical deep learning approach, we also test these data by the classical image super-resolution method using deep convolutional networks (SRCNN) \cite{DongLoy-324}.

\subsection{Training image 0 and image 1  separately}

\begin{figure}
\centering
	\subfloat[The training losses of image 0.]{\label{fig4-1}
        \includegraphics [scale=0.4]{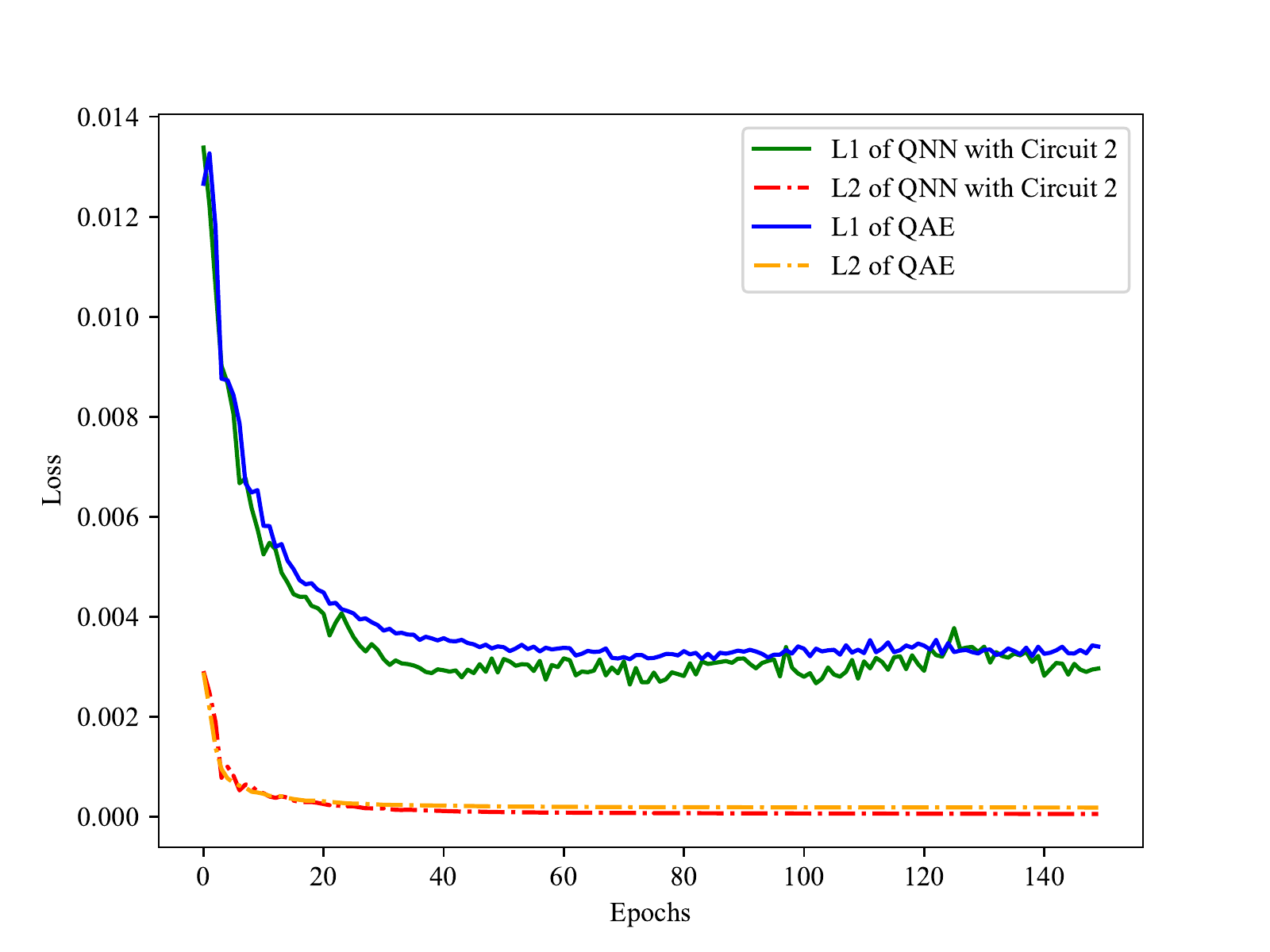}}
	\subfloat[The fidelities of testing sample of image 0.]{\label{fig4-2}
        \includegraphics [scale=0.4]{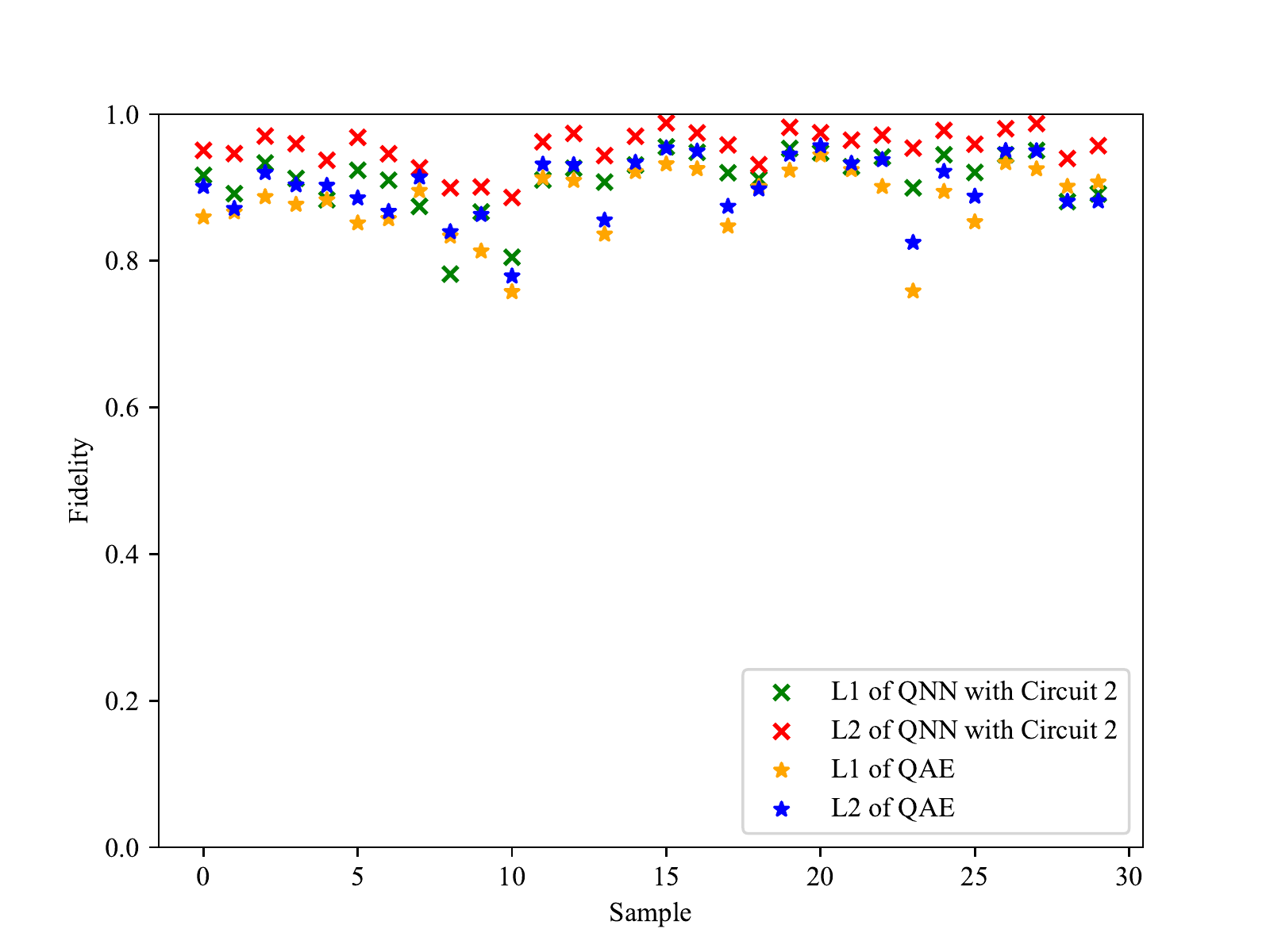}}\\
	\subfloat[The training losses of image 1.]{\label{fig4-3}
        \includegraphics [scale=0.4]{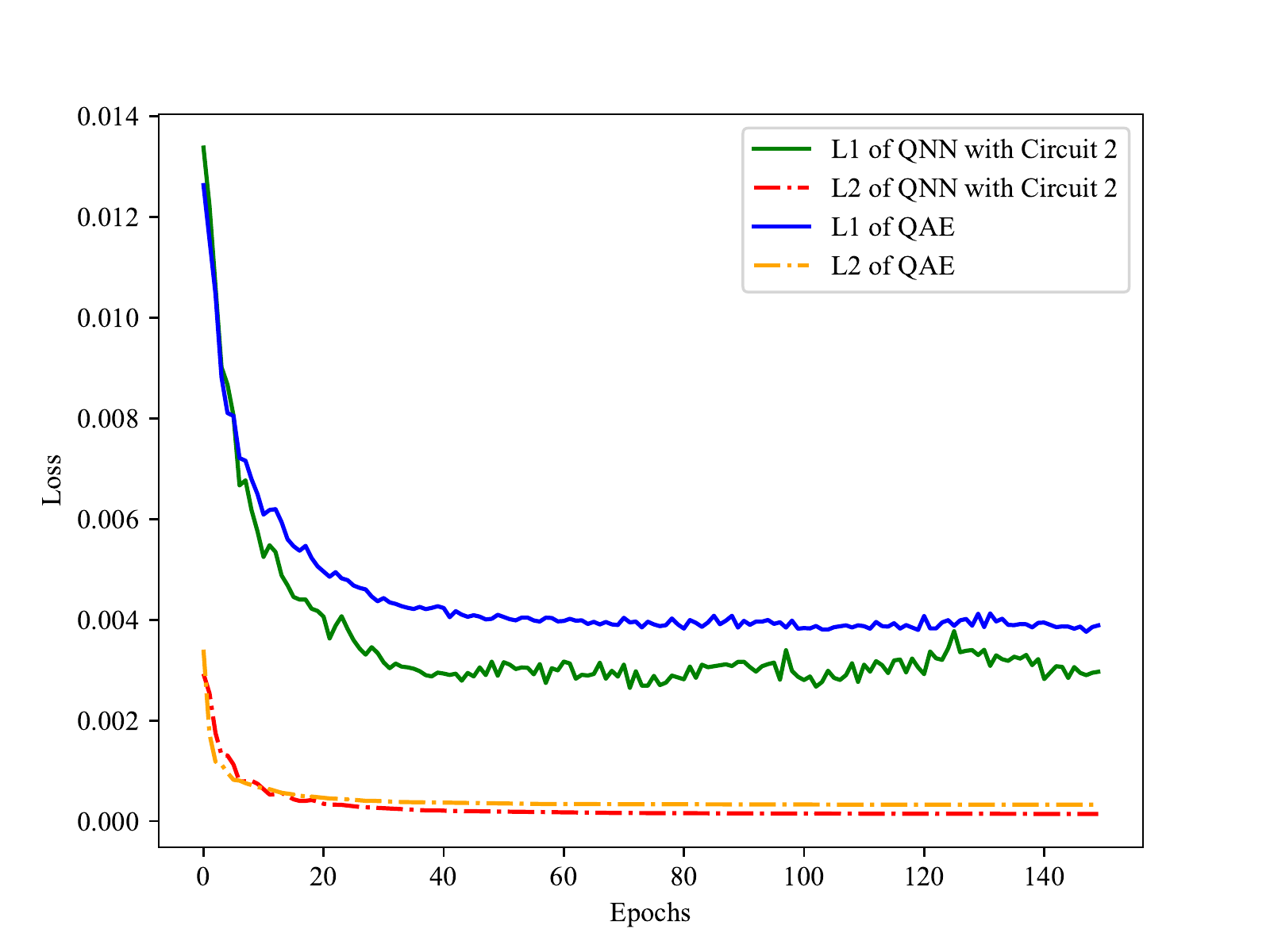}}
	\subfloat[The fidelities of testing sample of image 1.]{\label{fig4-4}
        \includegraphics [scale=0.4]{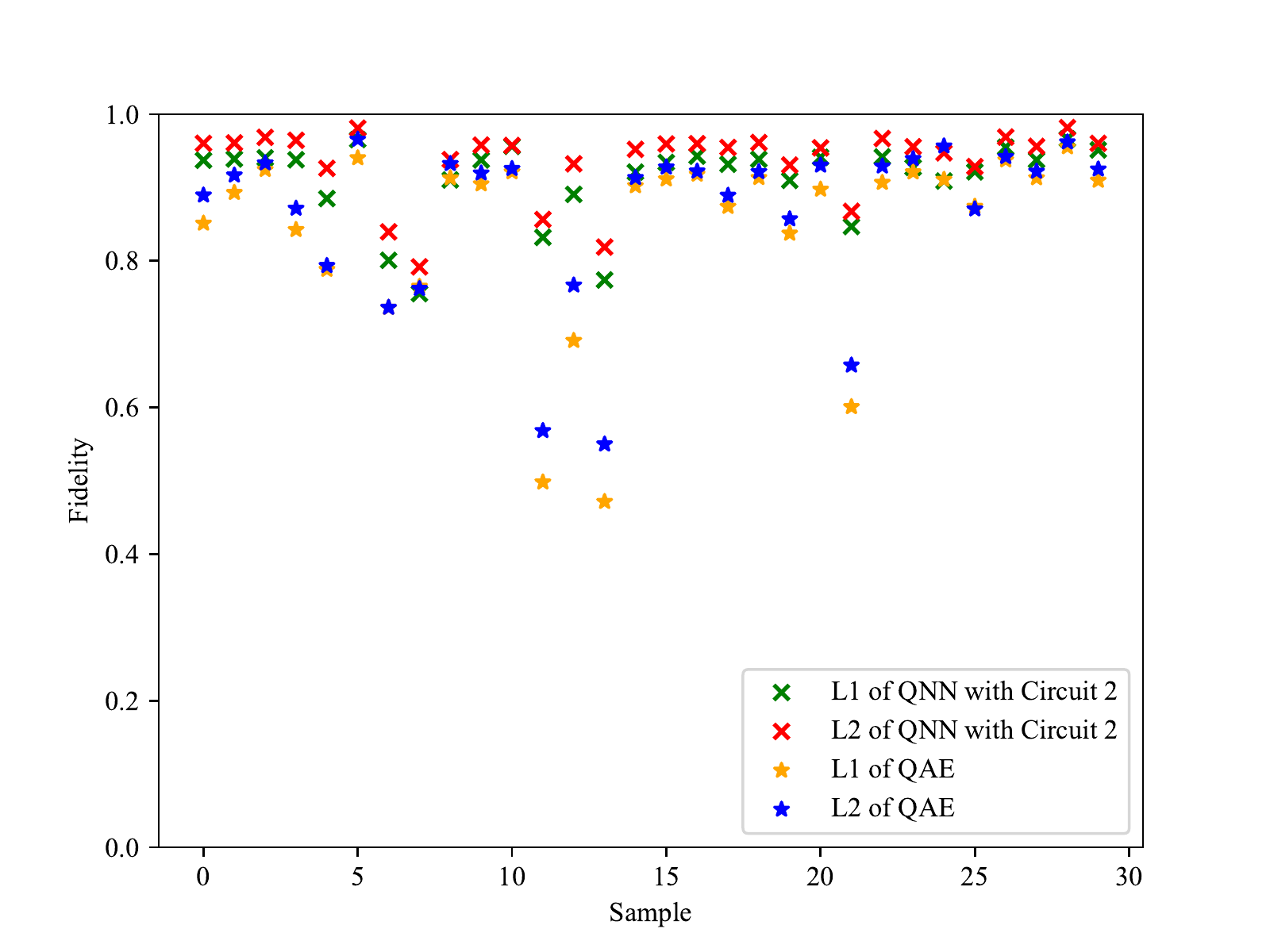}}
  \caption{Comparison of  $L_1$ and $L_2$ for training and testing of the QNN with Circuit 2 and the QAE by taking image 0 and image 1 as training sets separately.}
  \label{fig4}
\end{figure}

We show the simulation of QNNs and QAE by taking image 0 and image 1 as training sets, separately.
Firstly, we compare the difference of the QNN with Circuit 2 and the QAE for taking $L_1$ and $L_2$ as loss functions.
As is shown in Fig. \ref{fig4}, using the same optimizer and the same epoch, $L_1$ and $L_2$ are steadily decreasing with related fidelities close to 1. Besides,  $L_2$ performs better in terms of the loss convergence and corresponding fidelity.
Furthermore, the QNN with Circuit 2 using $L_2$ can obtain a higher average fidelity than the QAE.

\begin{figure}
\centering
	\subfloat[The training losses of image 0.]{\label{fig5-1}
        \includegraphics [scale=0.4]{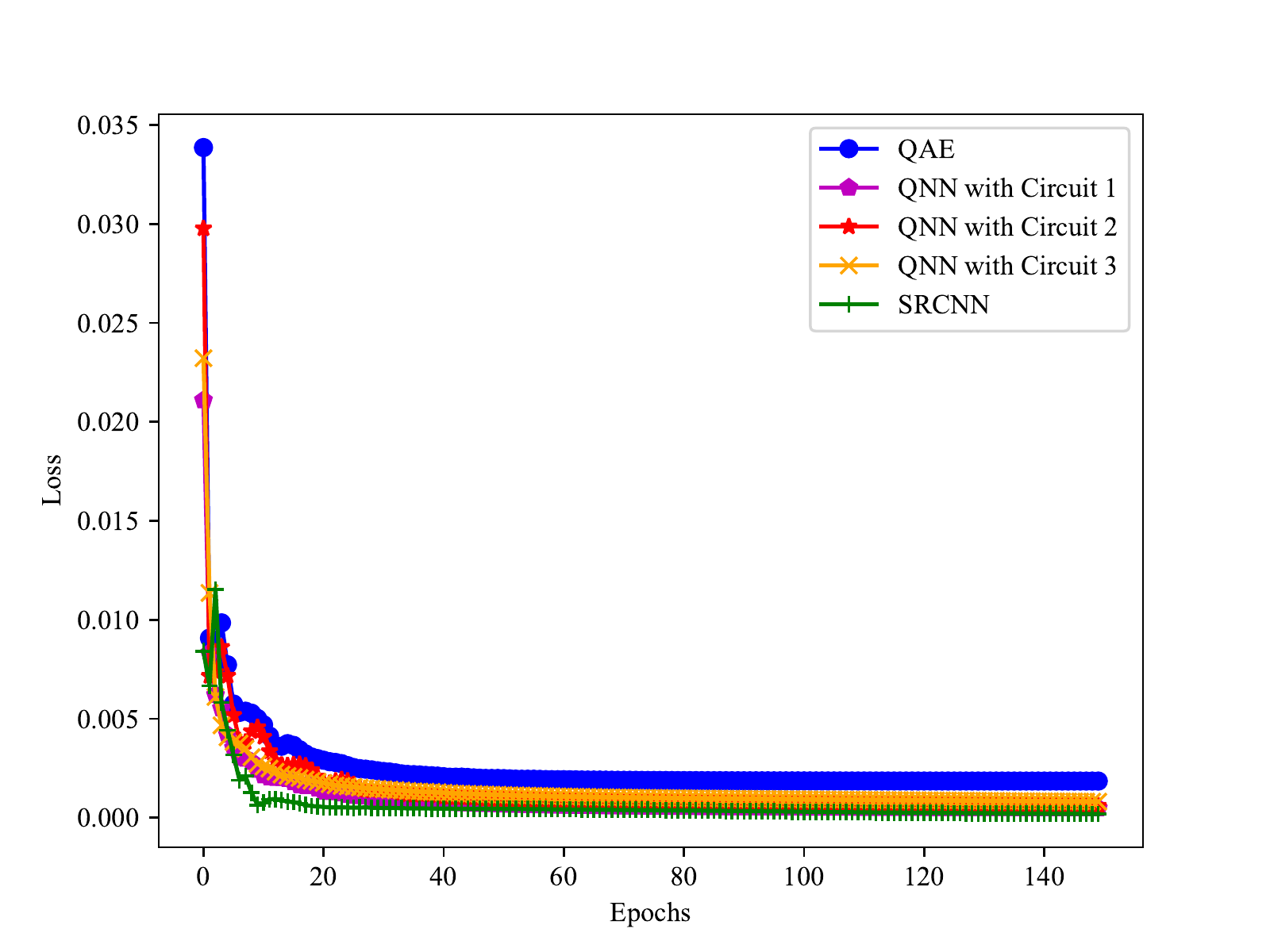}}
	\subfloat[The fidelities of testing sample of image 0.]{\label{fig5-2}
        \includegraphics [scale=0.4]{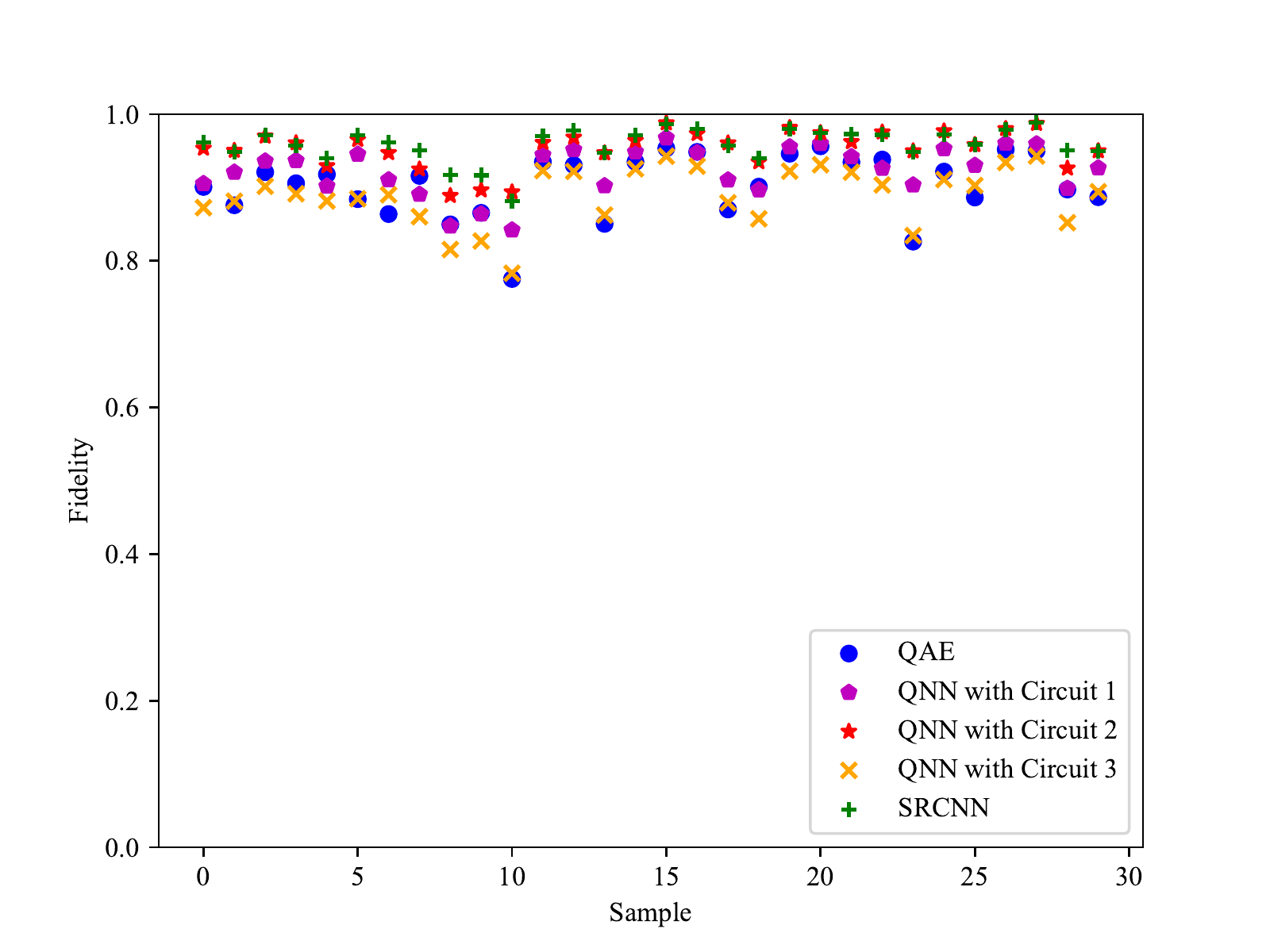}}\\
	\subfloat[The training losses of image 1.]{\label{fig5-3}
        \includegraphics [scale=0.4]{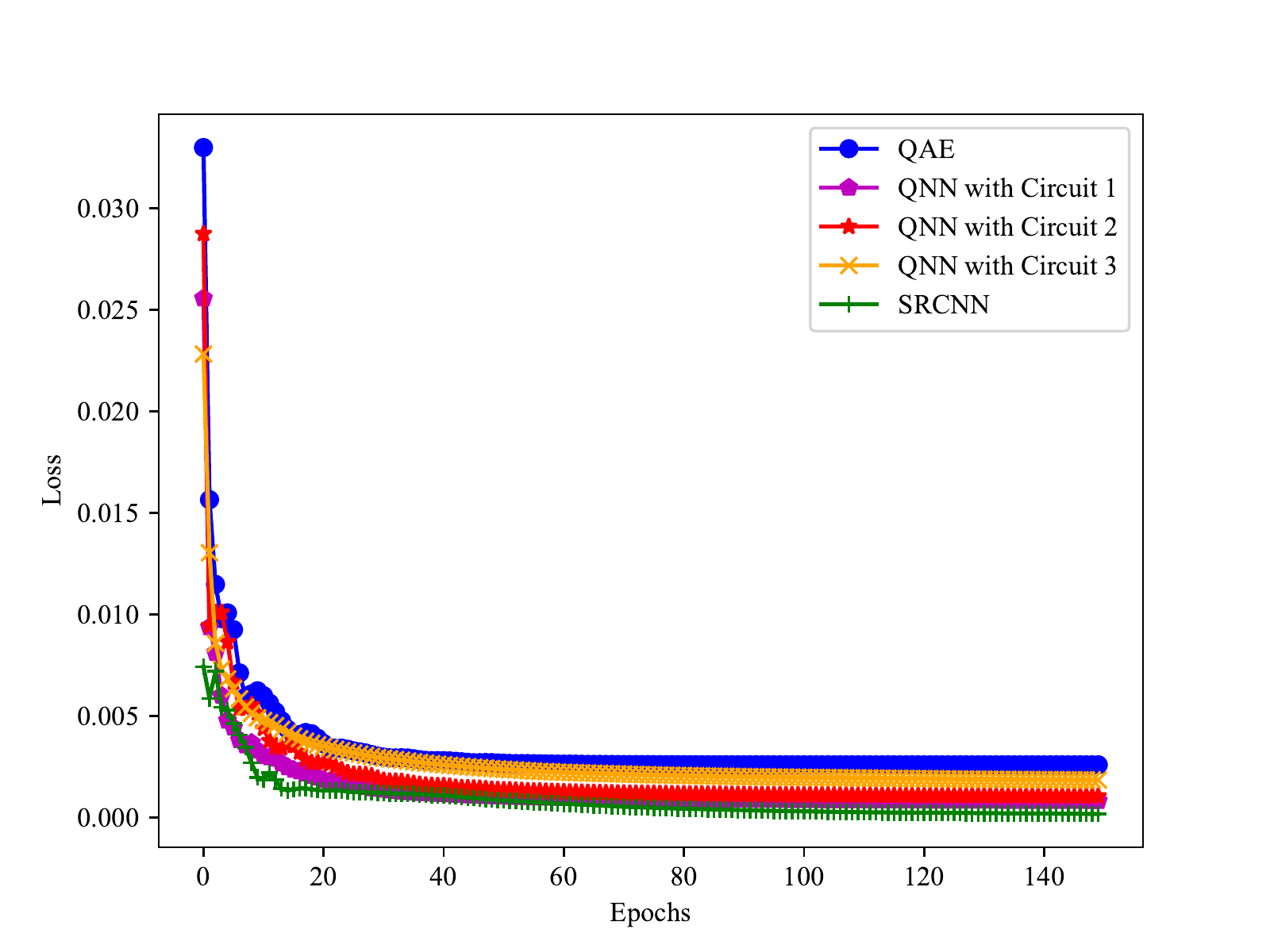}}
	\subfloat[The fidelities of testing sample of image 1.]{\label{fig5-4}
        \includegraphics [scale=0.4]{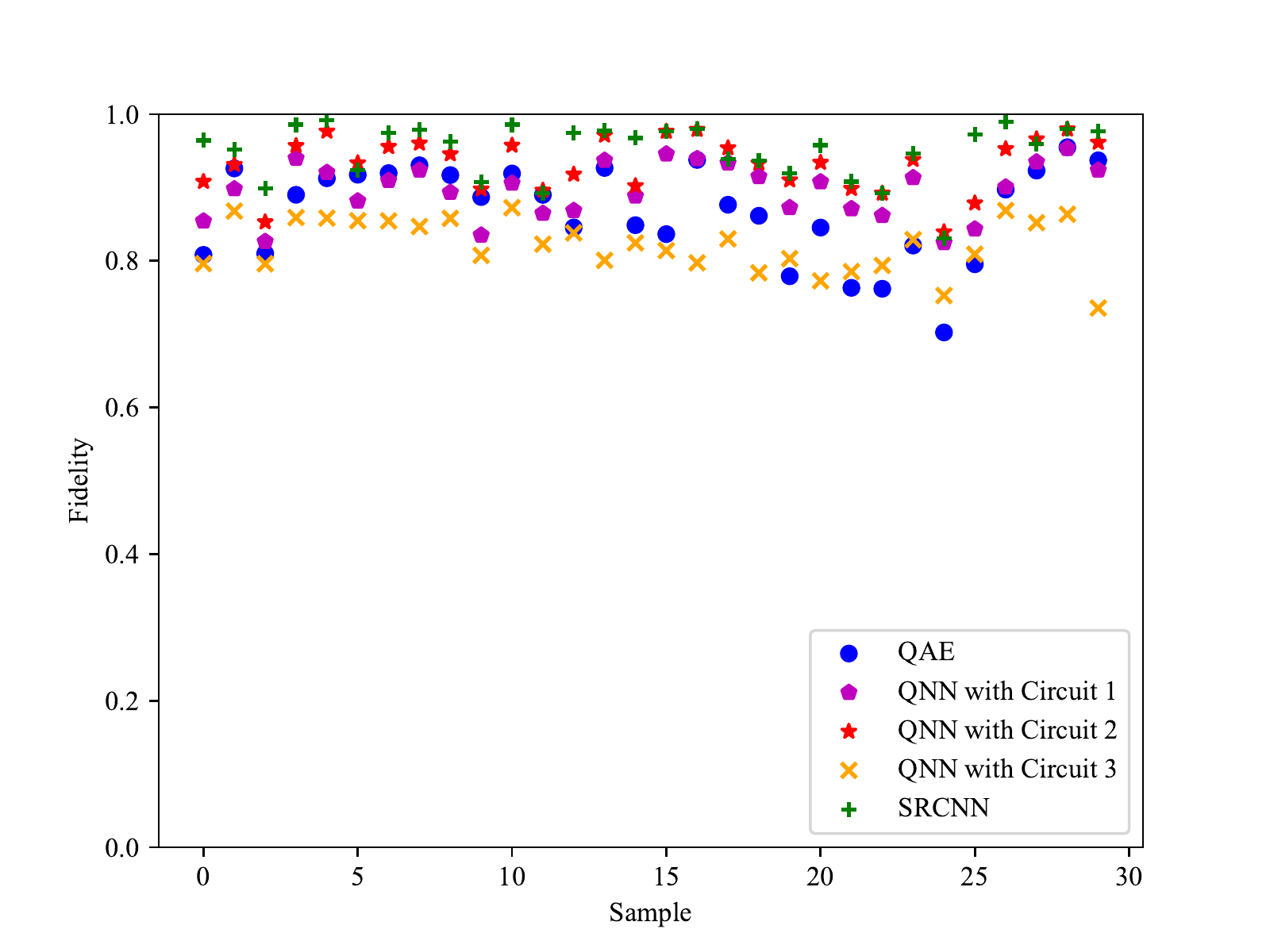}}
  \caption{The training and testing processes of QNNs, QAE, and SRCNN for taking image 0 and image 1 as training sets separately.}
  \label{fig5}
\end{figure}

\begin{figure}
\centering
	\subfloat[The image 0 set]{\label{fig6-1}
		\includegraphics [scale=0.27]{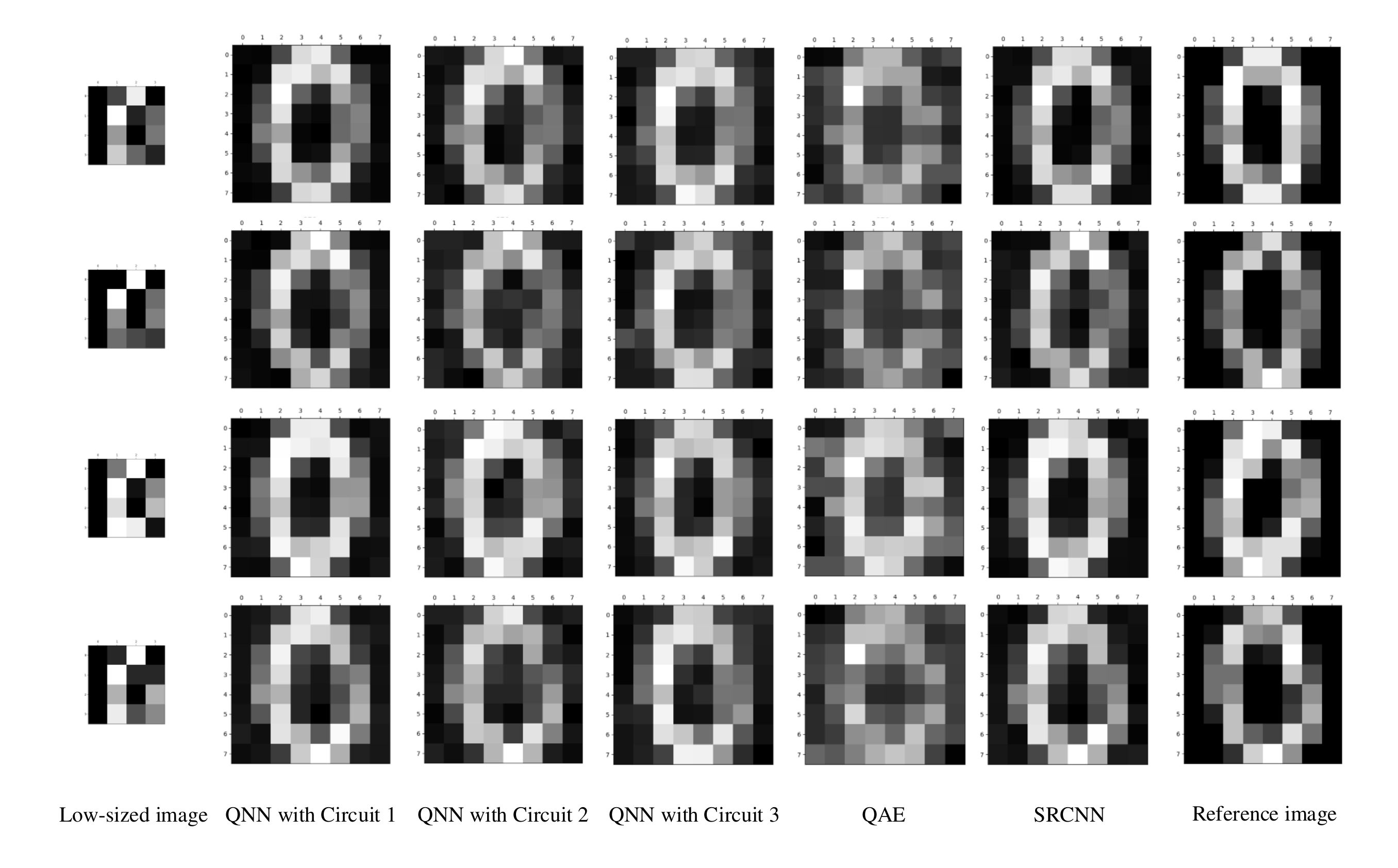}}	
	\subfloat[The image 1 set]{\label{fig6-2}
		\includegraphics [scale=0.27]{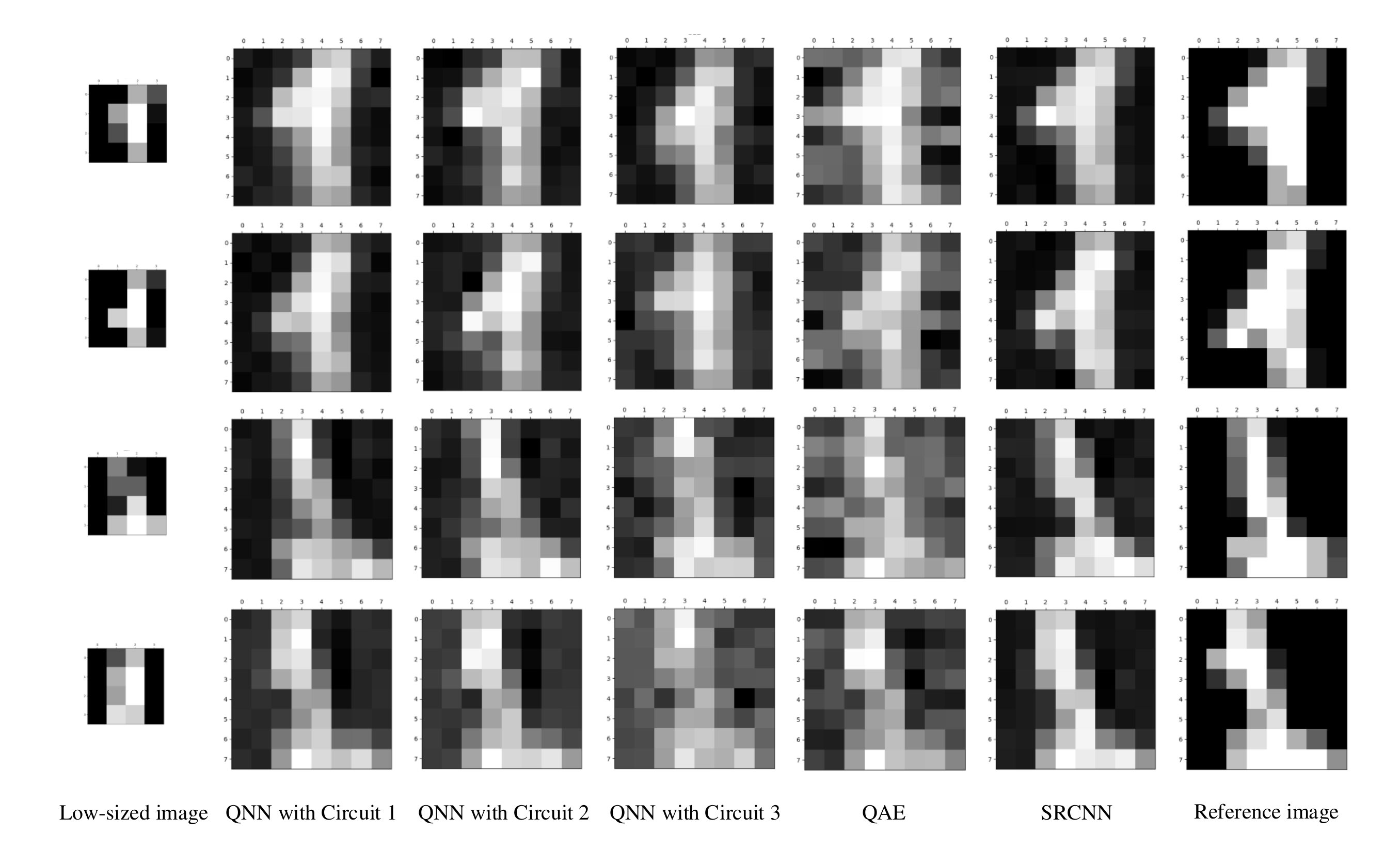}}
\caption{Reconstruction effects of QNNs with Circuit 1, Circuits 2, and Circuits 3, QAE, and SRCNN for taking image 0 and image 1 as training sets separately, where 1) the 1st column of each subfigure denotes the low-sized images; 2) the 2nd, 3rd, and 4th columns denote the reconstructed images by QNNs with Circuits 1, 2, and 3 respectively; 3) the 5th column denotes the reconstructed images by QAE; 4) the 6th column denotes the reconstructed images by SRCNN; 5) the last column denotes the reference images.}
\label{fig6}
\end{figure}

In the following, we will only show the results of $L_2$. The training losses and testing fidelities of QNNs with three Circuits, the QAE, and the SRCNN are shown in Fig. \ref{fig5}.
While images reconstruction effects of some digits are presented in Fig. \ref{fig6}, which show that three QNNs and the QAE can get relatively good data reconstruction effects.

\subsection{Training mixed image 0 \& 1 set}

\begin{figure}
\centering
	\subfloat[The training losses.]{\label{fig7-1}
		\includegraphics [scale=0.4]{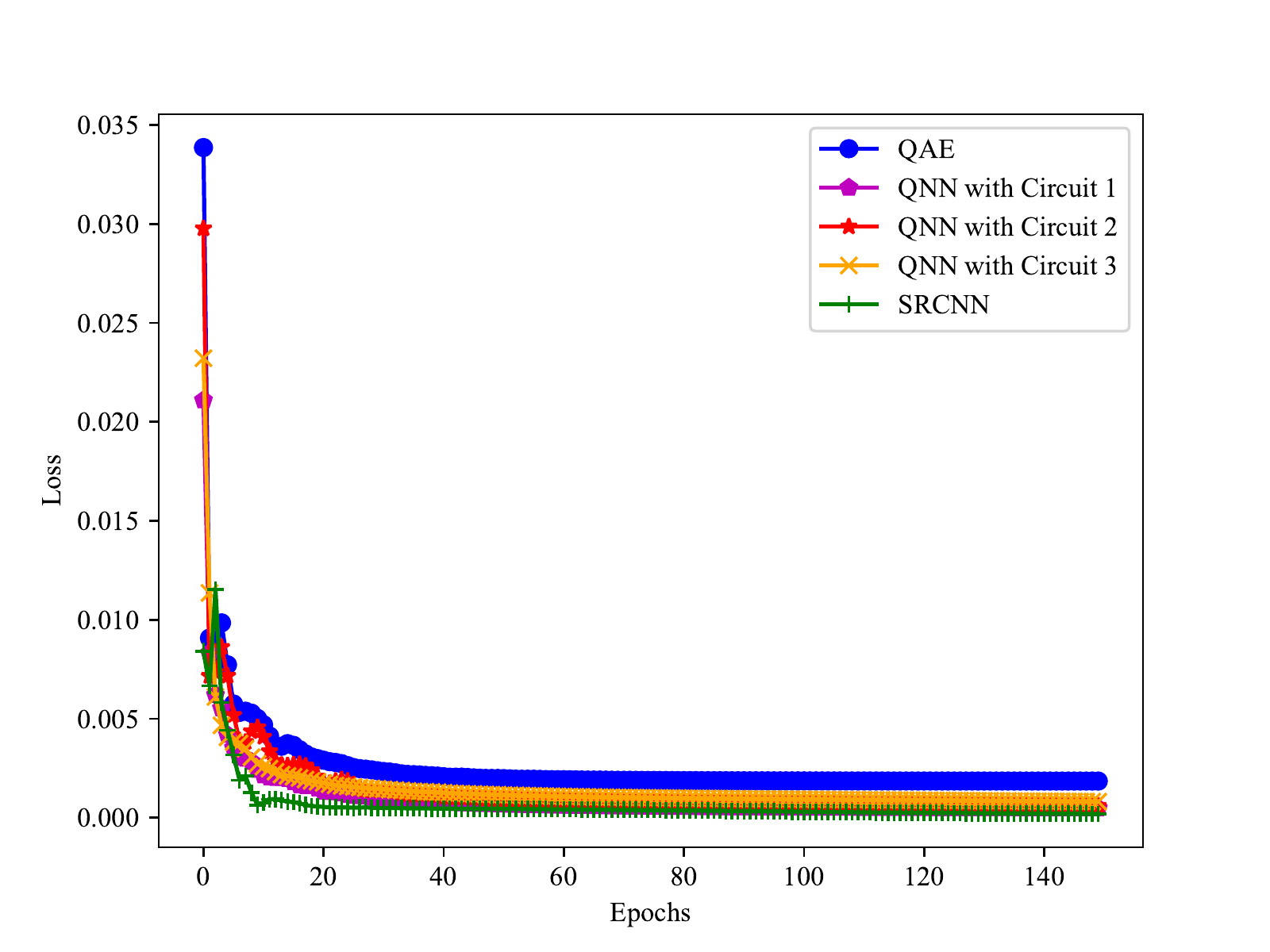}}
	\subfloat[The fidelities of testing sample.]{\label{fig7-2}
		\includegraphics [scale=0.4]{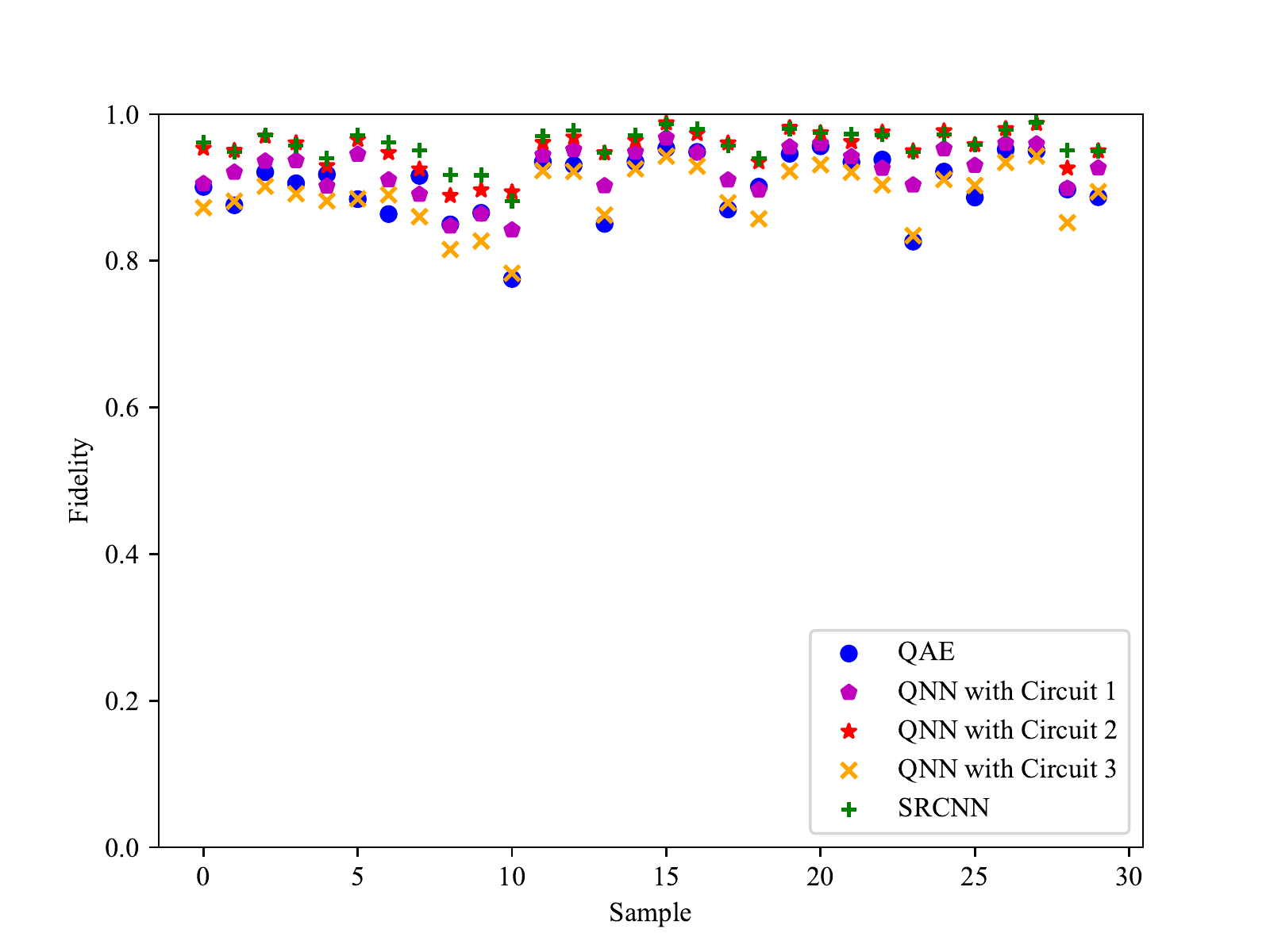}}
  \caption{The training and testing for mixed image 0 and 1 set.}
  \label{fig7}
\end{figure}

\begin{figure}
\centering
\includegraphics [scale=0.27]{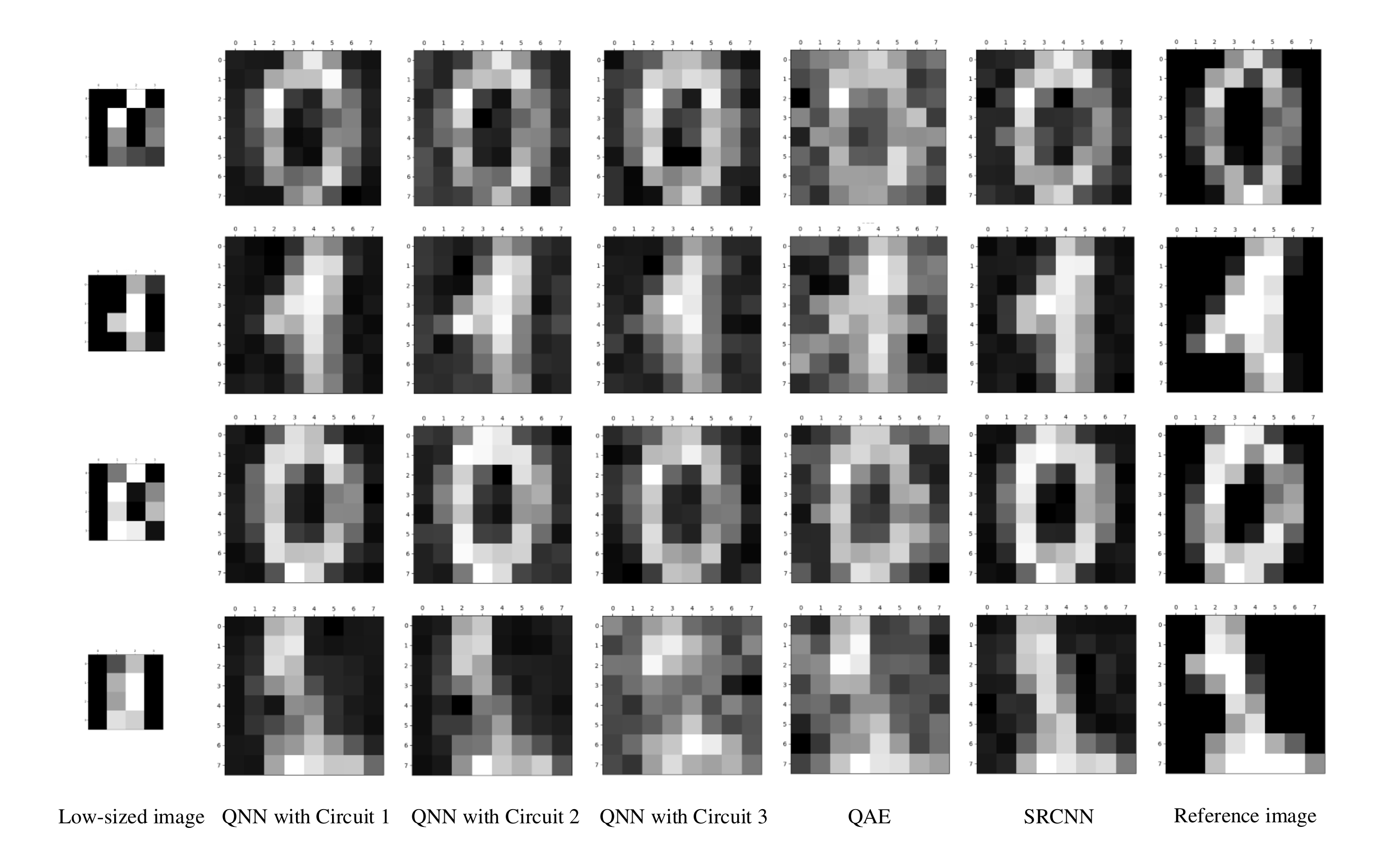}
  \caption{Reconstruction effects of QNNs, QAE, and SRCNN for taking mixed images 0 \& 1 as the training set.}
 \label{fig8}
\end{figure}

Our QNN-based and QAE-based frameworks can reconstruct data even for mixed training of different types of digits.
The training and testing processes of QNN and QAE taking mixed images 0 \& 1 as training set are shown in Fig. \ref{fig7}.
While reconstruction effects of some mixed images are shown in Fig. \ref{fig8}.

\subsection{Results}

Tab. \ref{tab1} shows the average $L_2$ losses, fidelities, and parameter numbers of QNNs, the QAE, and the classical SRCNN for testing the same set of sample, where the circuit depth of QNNs and QAE are set to 40, and the training epoch is set to 150, respectively. The QNN-based and QAE-based frameworks have shown good reconstruction effects regardless of whether the same type of sample or different types of mixed sample are used.
Our best quantum model is the QNN with Circuit 2, which uses less number of parameters than the Circuit 1 and QAE for achieving a better reconstruction effect. While the other three quantum models still get acceptable  reconstruction effects.
Besides, fidelities of the QNN with Circuit 2 are close to the classical SRCNN, which means QNNs can achieve nearly the same effect for image super-resolution as CNN but with much less number of parameters.

\begin{table}[!htb]
\begin{center}
\caption{Testing results of QNNs, QAE, and SRCNN.}
\label{tab1}
\begin{tabular}{lccccc}
\hline\noalign{\smallskip}
 & QNN-Circuit 1 & QNN-Circuit 2 & QNN-Circuit 3 & QAE & SRCNN\cite{DongLoy-324} \\
\noalign{\smallskip}\hline \noalign{\smallskip}
$\bar{L}_2$ / $\bar{F}$ of image 0   & 0.0011/0.923 & 0.0015/\underline{\textbf{0.953}} & 0.0016 / 0.889 &0.0030/0.903 & 0.0013/\textbf{0.959}\\
$\bar{L}_2$ / $\bar{F}$ of image 1    & 0.0015/0.896& 0.0021/\underline{\textbf{0.932}} & 0.0021 / 0.822   &0.0041/0.868 & 0.0015/\textbf{0.950}\\
$\bar{L}_2$ / $\bar{F}$ of image 0 \& 1  & 0.0020/0.883 & 0.0023/\underline{\textbf{0.925}}& 0.0023 / 0.830  & 0.0045/0.858 & 0.0017/\textbf{0.945}\\
\noalign{\smallskip}
Param \#  & 960 & 480 & \textbf{240} &482 &62529 \\
\noalign{\smallskip}
\noalign{\smallskip}\hline
\end{tabular}
\end{center}
\end{table}

\section{Discussions \& Conclusions}

Some recent researches are related to our works, where their comparisons are shown in Tab. \ref{tab2}.
On the one hand, QAE has been proposed to compress quantum or classical data \cite{RomeroOlson-104,Bravo-Prieto-203} for reducing data size.
While our task is to reconstruct large-sized data from low-sized ones. Our QAE-based framework is similar to the decompressing part of QAE in Refs. \cite{RomeroOlson-104,Bravo-Prieto-203}.
But we use a slightly different loss function for our training and testing.
On the other hand, QuGAN \cite{BracciaCaruso-44,Hellstem-47} has been used to generate artificial images. But their process is different from ours since they have different types of input for the models for different types of tasks.

\begin{table}[htp]
\caption{Comparisons of our work with Refs. \cite{RomeroOlson-104,Bravo-Prieto-203,SteinBaheri-51,RudolphToussaint-52}.}
\label{tab2}
\begin{center}
\begin{tabular}{lccc}
\hline\noalign{\smallskip}
  & Refs. \cite{RomeroOlson-104,Bravo-Prieto-203} & Refs. \cite{SteinBaheri-51,RudolphToussaint-52} & Ours \\
  \noalign{\smallskip}\hline \noalign{\smallskip}
Tasks & Data compression & Image generation & Data reconstruction \\
Networks & QAE & QuGAN & QNN \& QAE\\
\noalign{\smallskip}\hline
\end{tabular}
\end{center}
\end{table}

In our two frameworks, the selection of loss functions will have different results in the training and evaluation processes of QNNs and QAE. As is shown in Fig. \ref{fig4}, $L_2$ is more suitable than $L_1$ for the image reconstruction task. For comparing $L_2$ and $L_1$, fidelity is used to measure the differences between output states and reference states.
Besides, we also compare our quantum models with CNN, which shows that quantum models can achieve nearly the same result for solving classical problems, but require much less number of model parameters.

Besides, the selection of circuit structures (or ansatz) greatly affects the result of QNN and QAE for our data reconstruction task.
We use three circuits (Circuit 1, Circuit 2, and Circuit 3) to test our QNNs, where the QNN with Circuit 2 has the best data reconstruction effect in our experiments, while the QNNs with Circuit 1 and the QAE, although using more training parameters, have relatively low reconstruction fidelity. The QNNs with Circuit 3, which requires the least number of parameters, can still get an acceptable  reconstruction effect. It should be noted that other types of circuits are also possible be used on QNNs and QAE for achieving better results.

In conclusion, we have proposed two QML frameworks, i.e., the QNN-based one and the QNN-based one, to deal with the data reconstruction problem from low-sized data to large-sized ones.
To show the effects of the two frameworks, we simulate them by using MNIST handwritten digits as our datasets. PQCs are used to build QNNs and the QAE where learnable parameters can be optimized iteratively. Simulation results show that two frameworks are feasible for data reconstruction.
In the future, it is interesting to explore other types of QML frameworks for dealing with the data reconstruction problem.

\section*{Data availability statement}
The data that support the findings of this study are available from the corresponding author upon reasonable request.

\section*{Acknowledgements}
This project was supported by the National Natural Science Foundation of China (Grant No. 61601358).


\end{document}